\documentclass[final,5p,times]{article}

\usepackage{multirow,booktabs}

\usepackage{amsmath}

\usepackage{listings}
\usepackage{float}
\usepackage{rccol}
\usepackage[table]{xcolor}
\usepackage{algorithm2e}
\usepackage{wrapfig}
\usepackage{graphicx}
\usepackage{plain}
\usepackage{authblk}
\usepackage{amsmath}
\usepackage{comment}

\usepackage[normalem]{ulem}

\usepackage{xr}

\usepackage{setspace}
\usepackage[margin=0.7in]{geometry}
\usepackage{amssymb}
\usepackage{makecell}

\begin{document}
%\doublespacing
\noindent \large{\textbf{A structurally informed model for modulating functional connectivity}}
\\
\\
Andrew C. Murphy\textsuperscript{1,2}, Romain Duprat\textsuperscript{3}, Theodore D. Satterthwaite\textsuperscript{4,5}, Desmond J. Oathes\textsuperscript{3,6}, Dani S. Bassett\textsuperscript{1,4,7-10,*}
\\
\\
\textsuperscript{1}Department of Bioengineering, School of Engineering \& Applied Science, University of Pennsylvania, Philadelphia, PA 19104, USA
\\
\textsuperscript{2}Perelman School of Medicine, University of Pennsylvania, Philadelphia, PA 19104, USA
\\
\textsuperscript{3}Center for Neuromodulation in Depression and Stress, Department of Psychiatry, Perelman School of Medicine, University of Pennsylvania, Philadelphia, PA 19104, USA
\\
\textsuperscript{4}Department of Psychiatry, Perelman School of Medicine, University of Pennsylvania, Philadelphia, PA 19104, USA
\\
\textsuperscript{5}Penn Lifespan Informatics and Neuroimaging Center, Perelman School of Medicine, University of Pennsylvania, Philadelphia, PA 19104, USA
\\
\textsuperscript{6}{Penn Brain Science, Translation, Innovation, and Modulation Center, Department of Neurology, Perelman School of Medicine, University of Pennsylvania, Philadelphia, PA 19104, USA}
\\
\textsuperscript{7}Department of Physics \& Astronomy, College of Arts \& Sciences, University of Pennsylvania, Philadelphia, PA 19104, USA
\\
\textsuperscript{8}Department of Neurology, Perelman School of Medicine, University of Pennsylvania, Philadelphia, PA 19104, USA
\\
\textsuperscript{9}Department of Electrical \& Systems Engineering, School of Engineering \& Applied Science, University of Pennsylvania, Philadelphia, PA 19104, USA
\\
\textsuperscript{10}Santa Fe Institute, Santa Fe, NM 87501, USA
\\
\textsuperscript{*}To whom correspondence should be addressed: dsb@seas.upenn.edu

\clearpage
\section{Abstract}
Functional connectivity between brain regions tracks symptom severity in a wide variety of neuropsychiatric disorders. Transcranial magnetic stimulation (TMS) is a powerful tool to directly alter regional activity and indirectly alter functional connectivity. Predicting how the strength of a functional connection will change following neuromodulation with repetitive TMS (rTMS) is difficult, but would allow novel therapies that target functional connections to improve symptoms.
Here, we address this challenge by proposing a predictive model that explains how TMS-induced activation can change the strength of a functional connection. Cortical stimulation can result in spatially distributed functional connectivity changes across the cortex. Here, we focus on the frontoparietal (FPS) and default mode (DMS) systems given the importance of their functional connection in executive function and affective disorder pathology. We fit this model to neuroimaging data in 29 individuals who received a single round of acute rTMS to the frontal cortex and evaluated the functional connectivity between the FPS and DMS.
%\DB{Frontal cortical stimulation can result in spatially distributed functional connectivity changes across the cortex. Here, we focus on the frontoparietal and default mode systems given the importance of their functional connection in executive function and affective disorder pathology. }
%To assess generalizability, we divided the cortex into 400 discrete non-overlapping regions. 
For each individual, we measured the TMS-induced change in functional connectivity between the FPS and DMS (the \emph{functional connectivity network}), and the structural coupling between the stimulated area and the FPS and DMS (the \emph{structural context network}). We find that TMS-induced changes in functional connectivity are best predicted when the model accounts for white matter fibers from the stimulated area to the two systems. We find that the similarity between these two networks (structure-function coupling)---and therefore the predictability of the TMS-induced modulation---was highest when the structural context network contained a dense core of intraconnected regions, indicating that the stimulated area had ample access to an anatomical module. Further, we found that when the core of the structural context network overlapped with the FPS and DMS, we observed the greatest change in the strength of their functional connection. Broadly, our findings explain how the structural connectivity of a stimulated region modulates TMS-induced changes in the brain's functional network. In the future, efforts to account for such structural connectivity could improve predictions of TMS response, further informing the development of neuromodulation protocols for clinical translation.
%
%\clearpage
\newpage
\section{Introduction}
\begin{figure}[H]
%\centerline{\includegraphics[]{fig0-01.png}}
\centerline{\includegraphics[width=\textwidth]{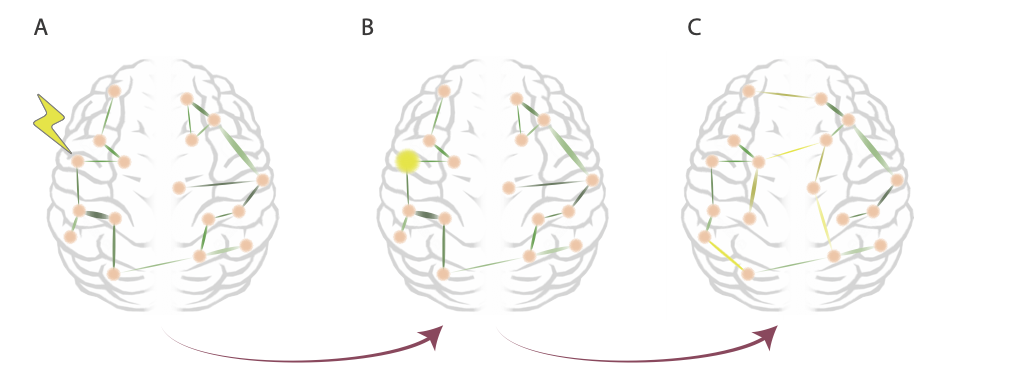}}
\caption{\textbf{The causal effect of TMS on the functional connectome.} We hypothesize the following causal pathway beginning with TMS application and resulting in the modulation of functional connections. \emph{(A)} First, TMS is applied to a specific region within the brain network. \emph{(B)} This application alters the activity of the stimulated region, which leads to \emph{(C)} changes in the functional connectome.} 
 \label{mod_design}
\end{figure}

While the advent of pharmacotherapies for neuropsychiatric disorders has been hugely beneficial \cite{mitchell1997effectiveness}, a sizable portion of patients do not experience the expected benefit from these treatments \cite{al2012treatment,chiliza2015rate}. The presence of these treatment refractory cases emphasizes the need for novel approaches to the diagnosis and treatment of neuropsychiatric disorders. Evidence from fMRI BOLD neuroimaging suggests that functional connectivity, a measure of the statistical dependence between regional time series \cite{noble2019decade}, is a useful correlate of clinical symptoms. Individual variation in functional connectivity tracks individual variation in symptom severity in Alzheimer's disease \cite{Fox2012c,schumacher2019dynamic}, depression \cite{tozzi2021reduced}, schizophrenia \cite{Lynall2013,sheffield2016cognition}, and attention-deficit/hyperactivity disorder (ADHD) \cite{Konrad2010}, as well as other neurological and psychiatric conditions. Beyond explaining individual variation, functional connectivity has also begun to show utility as a feature that can be modulated to design and personalize brain-based interventions. For example, functional connectivity can now be altered by applying transcranial magnetic stimulation (TMS) to a given brain region (Fig \ref{mod_design}). The approach of functional connectivity targeted TMS is currently used to treat depression \cite{Fox2012d,Liston2014,pridmore2018early,williams2021identifying}, obsessive-compulsive disorder \cite{Williams2021}, and post-traumatic stress disorder \cite{philip2019theta}. Central to the associated therapeutic model is the identification of a cortical stimulation point that will reliably alter the targeted functional connection. Although prior studies propose methods to choose stimulation points that reliably alter activity \cite{Wu2014,Fox2013a,Oathes2018}%, either directly \cite{Wu2014} or indirectly based on functional connections \cite{Fox2013a,Oathes2018}%,Halko2014,Volz2015}
, similar methods to choose stimulation points that reliably alter connectivity have not yet been developed. Such a development may be clinically useful for patients who have not responded to currently available treatments.

Here we hypothesize that a linear model, informed by the magnitude of TMS-induced regional \emph{activity} change in the stimulated region, will predict the resulting change in \emph{functional connection} strength between two other downstream regions. Further, we hypothesize that the addition of white matter connections to the model will further improve model accuracy in predicting changes in functional connection strengths. To develop a method to predict changes in functional connectivity from stimulation, we first seek a conceptual and technical model of how regional neuromodulatory stimulation maps to changes in functional connectivity. We begin by choosing a single large-scale functional coupling between two systems whose biological nature and cognitive relevance is well characterized. Specifically, we focus on the functional connection between the frontoparietal system (FPS) and the default mode system (DMS). In humans, the strength of the connection between the FPS and DMS is heavily dependent on task engagement: in a wide range of tasks, the FPS activates while the DMS deactivates in comparison to their respective activity levels at rest \cite{ClareKelly2008,Delaveau2017}. This inverse relationship in activity is thought to indicate that the two systems have opposing functional goals \cite{Boveroux2010,Hannawi2015}. The strength of the FPS-DMS functional connection tracks individual differences on executive function tasks \cite{ClareKelly2008}, such as the categorization of facial emotion \cite{Xin2014}, and several reasoning tasks \cite{Hearne2015}. Moreover, the functional connection between the FPS and the DMS can be empirically modulated by TMS \cite{Chen2013}, raising the possibility of altering this connection to treat deficits in executive function.

After choosing the functional connection of interest, we next built a model postulating a causal chain (Fig. \ref{mod_design}) from (i) acute TMS applied to a particular region, to (ii) a change in the activity of that stimulated region, and to (iii) subsequent downstream changes in functional connection strengths. Specifically, we propose a simple linear model that links the magnitude of acute TMS-induced change in regional activity to the magnitude of subsequent change in functional connectivity between the DMS and FPS. We inform the model with resting state fMRI imaging data collected before and after the application of a single acute round of rTMS in 29 individuals. Next, we ask whether the accuracy of this mapping improves when we use diffusion imaging data to measure the stimulated region's structural links to the DMS and FPS. Extending beyond this system-level investigation of a single connection, we move to a region-level investigation of all connections among 400 cortical areas. We hypothesize that the change of a functional connection between two downstream regions depends upon the stimulated area's \emph{structural context}: the strength of the stimulated area's structural connections to two regions that are functionally connected. We demonstrate a significant statistical correspondence between a stimulated region's structural context and subsequent acute rTMS-induced changes in functional connections throughout the brain. Finally, we show how this correspondence could be used to choose a cortical TMS target to induce the largest change in the strength of the DMS - FPS functional connection. Taken together, our findings demonstrate a strong relationship between functional network changes and the structural connections of a TMS target region, and suggest that accounting for those structural connections may improve predictions of TMS effects on functional brain networks.

\section{Materials and Methods}
\subsection{Imaging data acquisition and preprocessing}
Data was acquired in a two-day protocol: the first day included a baseline structural (T1) and fMRI resting state scan to enable localization of TMS targets. On a separate day, subjects underwent a pre-TMS fMRI resting state scan, followed by a round of acute TMS stimulation, followed by a post-TMS fMRI resting state scan. Further details of the study protocol are provided in the following sections.

\subsubsection{Data acquisition} \label{data_acquisition}
Subjects first underwent a baseline structural (T1) and fMRI resting scan to enable localization of TMS targets. On a separate day, subjects underwent a pre-TMS resting scan as described in more detail above. Immediately following the resting state scan, subjects received a round of intermittent theta burst (iTBS) TMS, which consisted of 50 Hz burst pulses of 3 stimulations each. The sequence was repeated at 5 Hz in 20 blocks of 10 sequences, amounting to 600 pulses (1800 stimulations) total over a time period of approximately 4 minutes. Approximately 7 minutes following neuromodulation (during which interleaved TMS/fMRI with non-neuromodulatory single TMS pulses were delivered; data reported elsewhere \cite{oathes2021resting}), subjects underwent a post-TMS resting state scan.

Resting state fMRI data were collected locally from 29 individuals with mean age (std)  = 30 (8) years; 15 were female and 14 were male. The MRI data were acquired on a 3 Tesla Siemens Prisma scanner. For baseline scans (structural T1w and resting state) a standard 32 channel head coil was used (Erlangen, Germany). The two baseline resting state fMRI scans were acquired with opposite phase encoding directions A $\rightarrow$ P and P $\rightarrow$ A (TR=800ms, TE=37ms, FA=52$^{\circ}$, FOV=208mm, $2\times2\times2$ mm voxels, 72 interleaved axial slices with no gap, 420 measurements). Subjects were instructed to keep their eyes open and remain as still as possible while attending to a central fixation cross. Any observed absolute movement exceeding 2mm during acquisition lead to a rerun of the sequence to promote data quality. Participants who were not able to provide low motion scans were excluded from the following analyses. Two resting state scans were collected per subject: one pre-rTMS scan, and one post-rTMS scan \cite{Oathes2018}. The resting state scans and acute TMS were separated by approximately 6 minutes. Structural high-resolution multi-echo T1-weighted MPR images were also collected for each of the subjects using the following parameters: TR=2400ms, TI=1060ms, TE= 2.24ms, FA=8$^{\circ}$, $0.8\times 0.8\times 0.8$mm voxels, FOV= 256mm, PAT mode GRAPPA, 208 slices. 

The TMS dataset did not include diffusion imaging, and therefore we instead chose 100 randomly selected subjects from the Human Connectome Project (HCP) S900 release \cite{VanEssen2013}. Consistent with prior work \cite{Betzel2019}, and for reasons described below, we computed a single group-averaged diffusion MRI network from the HCP as a representative diffusion network, also termed a \emph{wiring diagram} \cite{siddiqi2022causal}. Participants included 61 females and 39 males, with a mean age (std) = 28.7 (3.65) years. All analyses were performed in accordance with the ethical regulations of the WU-Minn HCP Consortium Open Access Data Use Terms, and were approved by the IRB of the University of Pennsylvania. All participants provided informed consent in writing. Diffusion tensor images were collected with the following parameters: TR = 5520 ms, TE = 89.5 ms, flip angle = 78$^{\circ}$, refocusing flip angle = 160$^{\circ}$, FOV = 210$\times$180, matrix = 168$\times$144, slice thickness = 1.25 mm, number of slices = 111 (1.25 mm isotropic), multiband factor = 3, echo spacing = 0.78 ms, $b$-values = 1000, 2000, and 3000 s/mm2 \cite{Oathes2018}.

\subsubsection{Processing of functional magnetic resonance imaging data}
All of the fMRI data were preprocessed through the XCP Engine modular pipeline developed at the PennLINC, University of Pennsylvania \cite{tustison2014advanced}. First the signal was temporally bandpass filtered using a fast Fourier transform implemented with AFNI's 3DBandPass, with a pass range of 0.01--0.08 Hz. Frames with greater than 0.5 mm framewise displacement were removed as outliers. The signals were spatially smoothed (5mm FWHM) using FSL's SUSAN. Motion artifact was modeled as a linear combination of 36 timeseries: 6 realignment parameters estimated during preprocessing (x-, y-, and z-translations, roll, pitch, yaw) \cite{chen2007brain}, the mean timeseries in deep white matter ($1 + 6 = 7$), the mean timeseries in deep cerebrospinal fluid ($1 + 7 = 8$), the mean signal across the entire brain (global signal) ($1 + 8 = 9$), the first temporal derivative of the above timeseries ($9 \times 2 = 18$), and quadratic expansions of the above timeseries (18 $\times$ 2 = 36). These nuisance timeseries were removed from the BOLD signal using a general linear model to estimate residuals. When combined with the voxelwise despiking procedure completed during preprocessing, this confound regression approach is among the most effective methods for attenuating motion-related noise \cite{hallquist2013nuisance}. %, and global signal. %\DB{Why were the pre- and post-TMS reseting state data preprocessed in a different way than the baseline? I think we need to reprocess to make sure that all resting state fMRI data were treated in exactly the same way.}

\subsubsection{Processing of diffusion MRI data}

For the diffusion tensor imaging (DTI), the Human Connectome Project team applied an intensity normalization across runs, and also applied the TOPUP algorithm for EPI distortion correction \cite{Andersson2003}, the EDDY algorithm for eddy current and motion correction \cite{Andersson2016}, and the gradient nonlinearity correction, before then calculating the gradient $b$-value/$b$-vector deviation, and registering the mean $b$0 to the native volume T1w with FLIRT \cite{jenkinson2002improved}. BBR+bbregister and transformation of diffusion data, gradient deviation, and gradient directions to 1.25 mm structural space were also applied. The brain mask is based on the FreeSurfer segmentation. The BedpostX (Bayesian Estimation of Diffusion Parameters Obtained using Sampling Techniques) output was then calculated \cite{behrens2007probabilistic}, where the `X' stands for modeling crossing fibers. Markov Chain Monte Carlo sampling was used to build probability distributions on diffusion parameters at each voxel \cite{behrens2007probabilistic}. The process creates all of the files necessary to run probabilistic tractography. Using the Freesurfer recon-all data computed by the Human Connectome Project, the fsaverage5 space cortical parcellation was registered to the subject's native cortical white matter surface and then transformed to the subject's native diffusion volume space \cite{fischl2004automatically}. From these data, we derived seeds and targets for probabilistic tractography, which we ran with the FSL probtrackx2 algorithm using 1000 streams initiated from each voxel in a given parcel \cite{Murphy2019}. %\DB{We should add a citation to each sentence to justify the use of that approach.} 

\subsubsection{Whole brain functional parcellation}

We parcellated the brain into 400 discrete and non-overlapping cortical regions of interest using the Schaefer atlas \cite{Schaefer2017}. Of course, other functionally defined atlases exist, but they are less ideal for our purposes for several reasons: the Power atlas \cite{Power2011} does not provide full cortical coverage, and the Gordon \cite{Gordon2016a} and Brainnetome \cite{Fan2016a} atlases are not accompanied by well-validated cognitive network assignments.
%\DB{Can we state why we chose the Schaefer atlas? What is its strongest point in comparison to other atlases? No atlas is perfect, but it is nice to give a reason for why you chose this one.} 
The Schaefer atlas provides an assignment of each region to one of 17 putative cognitive systems: two visual, two somatomotor, two dorsal attention, two salience/ventral attention, one limbic, three frontoparietal, three default mode, and one temporo-parietal system. To ensure that the granularity of the data was consistent with the granularity of our hypotheses, and in accordance with our prior work \cite{Murphy2019}, we collapsed these 17 systems into 8 systems by combining individual systems that belonged to the same cognitive system; that is, we combined the two visual systems into a single system, the two somatomotor systems into a single system, the two dorsal attention systems into a single system, the two salience systems into a single system, the three frontoparietal systems into a single system, and the three default mode systems into a single system.

\subsection{Analysis of functional magnetic resonance imaging data}

\subsubsection{Estimation of functional connectivity matrices and regional activity}

We used the preprocessed resting state data to construct functional connectivity matrices that reflect pairwise interactions between systems or regions. Specifically, we extracted processed mean time series from each of the 400 regions in the Schaefer atlas. Consistent with much of the neuroimaging literature, we estimated the functional connectivity between regional time series using the Pearson correlation coefficient due to its interpretability \cite{Zalesky2012a}. We collated these estimates into a single $400 \times 400$ connectivity matrix, $C_f$, which we then treated as a formal network model \cite{Bassett2018}. In this network, regions were represented by network nodes, and functional connections were represented by weighted edges, where the weight of the edge between node $i$ and node $j$ was given by the Pearson correlation coefficient between the time series of region $i$ and the time series of region $j$. Finally, we separately averaged (i) the edge weights existing within a given cognitive system, and (ii) the edge weights existing between two systems; this process allowed us to construct a system-by-system connectivity matrix. To calculate change in regional BOLD activity, we first calculated activity as the statistical variance of the BOLD timeseries in accordance with our prior work \cite{Murphy2019}. We then calculated the absolute difference in the root-mean-square between the pre- and post-TMS time series.

\subsection{Analysis of diffusion tensor imaging data}

\subsubsection{Estimation of structural connectivity matrices}

After performing probabilistic tractography, we applied the same 400-region Schaefer atlas and calculated the proportion of streamlines seeded in a voxel in one region that reached another region \cite{Baum2018,Cao2013,Johansen-Berg2005}. 
We chose to use the proportion of streamlines to represent structural connectivity \cite{Baum2018} due to the inhomogeneity of the region sizes. This method allowed us to account for larger regions being more densely connected trivially due to their larger size. %\DB{I am not sure I am completely following this argument -- Can we cite a paper that says that for atlases with inhomogeneous region sizes, streamlines are best and why?}. 
We collated all inter-regional estimates of structural connectivity into a single $400 \times 400$ connectivity matrix, $C_{s}$, which we then treated as a formal network model of brain structure \cite{Bassett2018}. In this structural network, regions were represented by network nodes, and structural connections were represented by weighted edges, where the weight of the edge between node $i$ and node $j$ was given by the proportion of streamlines seeded at region $i$ that reach region $j$. Finally, we separately considered (i) the edge weights existing within a given system, and (ii) the edge weights existing between two systems; this process allowed us to construct a system-by-system connectivity matrix akin to the one that we constructed from functional data. All structural matrices were normalized by the total weight of all connections \cite{Baum2017}.

\subsubsection{Calculation of structural context}

\begin{figure}[H]
\centerline{\includegraphics[]{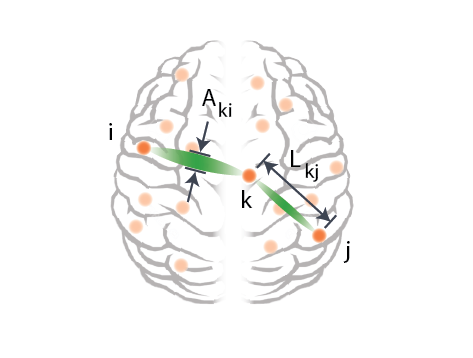}}
\caption{\textbf{Calculation of structural context.} To model white matter as wires, we calculated the length of a white matter wire as the Euclidean distance between white matter endpoints \emph{k} and \emph{i} within the brain ($L_{ki}$). We estimated the wire's cross-sectional area as the number of white matter fibers connecting the white matter endpoints \emph{k} and \emph{i} in the brain ($A_{ki}$), calculated from diffusion imaging.}
 \label{wiring}
\end{figure}

We posited that structural connections between systems would play an important role in the functional coupling between the frontoparietal and default mode systems. Specifically, we hypothesized that the formal nature of that role was one of boundary controllability \cite{Pasqualetti2014}. Boundary control is a metric predicated upon the notion that the topological location of a region within a network partially governs that region's influence on the function of modules or communities in the network \cite{Betzel2016a,Gu2015a}. Intuitively, if region \emph{k} has strong structural connections to regions \emph{i} and \emph{j}, then the (TMS-induced) activity of region \emph{k} influences the functional connection between regions \emph{i} and \emph{j}. To quantify this influence, we defined the \emph{structural context} (SC), which characterizes the position of a region ($k$) in relation to two other regions ($i$ and $j$). To formulate the structural context, we drew on concepts developed for neuronal cable theory. In modeling axons, neuronal cable theory assumes that resistance across axonal walls is much higher than along the core of the axon; this assumption is commonly referred to as the `core conductor' concept \cite{Holmes2013} and allows axons to be treated as wires within circuits \cite{hopfield1986computing}. Drawing on this framework, we formulate SC as the current flowing through a series of axonal wires from region \emph{i} to region \emph{k}, and then from region \emph{k} to region \emph{j} (Fig. \ref{wiring}).

%\begin{equation}
%\textup{SC}^{(k)}_{ij}=  \textup{B}_{ki} \times \textup{B}_{kj} \times \min \left(\textup{B}_{ki},\textup{B}_{kj} \right),
%\label{eq1}
%\end{equation}
%where $k$ is the stimulated region, and $\textup{SC}^{(k)}_{ij}$ quantifies the degree to which activity at $k$ may modulate the strength of the functional connection between regions $i$ and $j$. The matrix $\textup{B}$ is the measured $400 \times 400$ structural network, and the element $\textup{B}_{ki}$ is the strength of the structural coupling between regions $k$ and $i$. 
To calculate the current, we began by modeling the connections as two wires in series. Each of the wires has a resistance, with the total resistance being given by the following sum:
\begin{equation}
R_{ij} = R_{ki}+R_{kj}.
\label{eq1}
\end{equation}
The resistance of a wire between region \emph{k} and region \emph{i} is:
\begin{equation}
%R_{ki} = \rho \times \frac{L_{ki}}{A_{ki}} \propto \frac{L_{ki}}{A_{ki}}
R_{ki}  \propto \frac{L_{ki}}{A_{ki}},
\label{eq2}
\end{equation}
where $L_{ki}$ is the length of the connection, and $A_{ki}$ is the cross-sectional area of the wire, which we estimated as proportional to the number of white matter streamlines connecting regions $i$ and $k$, calculated from diffusion imaging (Fig. \ref{wiring}). The length of the connection is equal to the Euclidean distance between the two regions \emph{i} and \emph{j}. Making the above substitutions gives:
\begin{equation}
R_{ki} \propto \frac{L_{ki}}{B_{ki}},
\label{eq3}
\end{equation}
where \emph{B} is the $400 \times 400$ structural connectivity matrix, and $B_{ki}$ is the weight of the structural connection between regions \emph{k} and \emph{i}. Substituting Eq. \ref{eq3} into Eq. \ref{eq1} gives:
\begin{equation}
R_{ij} = R_{ki}+R_{kj} \propto \frac{L_{ki}B_{kj}+L_{kj}B_{ki}}{B_{ki}B_{kj}}.
\end{equation}
Finally, making use of Ohm's law, 
\begin{equation}
I \propto \frac{1}{R},
\end{equation}
the final equation for \emph{SC} is:
\begin{equation}
SC^{(k)}_{ij} = I \propto \frac{B_{ki}B_{kj}}{L_{ki}B_{kj}+L_{kj}B_{ki}}.
\label{sc}
\end{equation}
This equation is applied for all $i,j \in [1,N ]$ to generate a \emph{structural context} network of size $N \times N$ for any region \emph{k} (Fig. \ref{c4_fig0} D).
\subsubsection{Assessment of core-periphery structure}
\begin{figure}[H]
\centerline{\includegraphics[width=\textwidth]{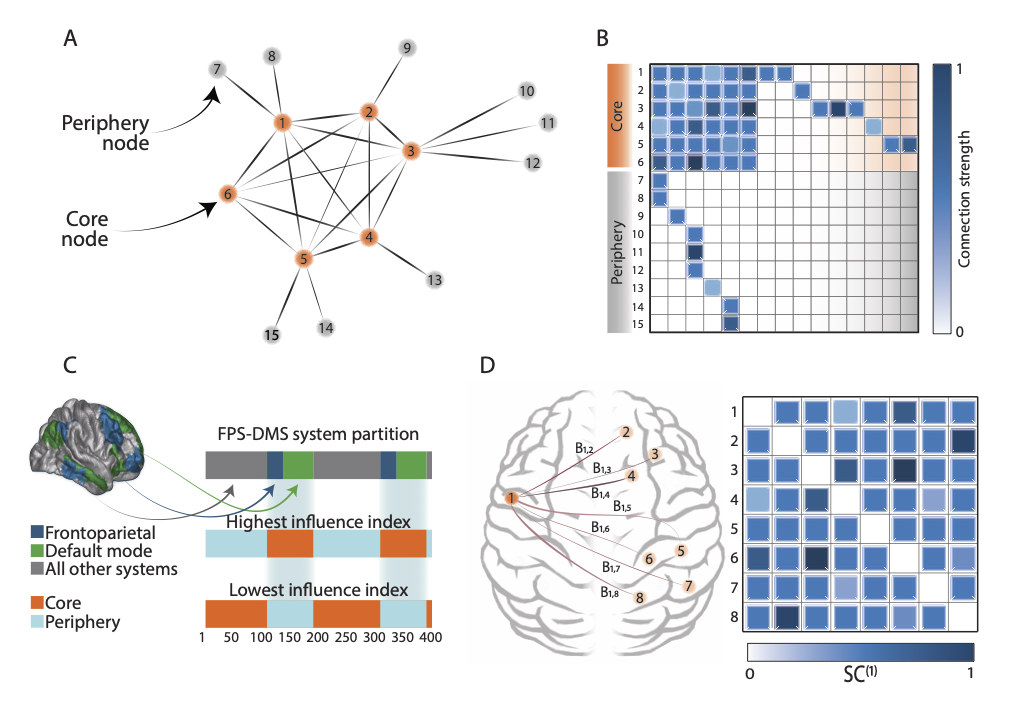}}
\caption{\textbf{Schematic of analytic strategy.}  $(A)$ A network with core-periphery structure consists of a dense core, where core nodes tend to connect with other core nodes, and a weakly connected periphery, where periphery nodes tend to only connect with core nodes and not with each other. Here, core-periphery structure is visualized by a spring embedding of a network where nodes that are connected to one another are placed near to one another and nodes that are not connected to one another are pushed away from one another. \emph{(B)} The same network can be visualized as an adjacency matrix, highlighting the densely-connected core in the upper left quadrent. $(C)$ The calculation of the influence index: The top bar shows the joint frontoparietal and default mode partition both represented on the brain surface and as a vector. This joint partition is composed of all non-gray regions. The influence index quantifies the similarity between the system partition and a core-periphery partition. A core-periphery partition with more overlap (middle bar) with the system partition has a high influence index, whereas a core-periphery partition with less overlap (bottom bar) with the system partition has a low influence index. $(D)$ A representation of the structural context matrix of a region within a network: the strength of the structural connections from region 1 to all other regions is calculated, and Eq. \ref{sc} is applied in a pairwise fashion for all regions to generate a full structural context matrix for region 1.}
 \label{c4_fig0}
\end{figure}

%\DB{Can we start this paragraph with the question we were hoping to answer or the hypothesis we wished to test, before describing the method we used to do so? Remind us why we need/want to assess core-periphery structure?}  
We hypothesized that a correspondence may exist between the architecture of a stimulated region’s structural context network and subsequent changes in functional connections throughout the brain in response to neuromodulation with acute rTMS. We anticipated that the nature of that correspondence would be related to the brain's core-periphery organization (Fig. \ref{c4_fig0} A-B), which is a feature of many networked systems in biology. Specifically, we hypothesized that the degree of core-periphery organization of a structural context network would correspond to subsequent acute rTMS-induced changes in functional connections within the brain. To quantify the presence and strength of core-periphery organization in the structural networks, we used a previously proposed method that is also implemented in the Brain Connectivity Toolbox \cite{Rubinov2015,Rubinov2010}. Briefly, this toolbox implements an algorithm to efficiently partition the network into core nodes and periphery nodes such that a core-ness statistic is maximized. For our adjusted structural network $\textup{B}'$ (see below), the core-ness statistic \emph{q} is:
\begin{equation}
q = \sum_{i,j \in \textup{R}_c} B'_{ij} - \sum_{i,j \in \textup{R}_p} B'_{ij}, \label{eq_coreness}
\end{equation}
where $\textup{R}_c$ is the set of core regions, and $\textup{R}_p$ is the set of periphery regions. This algorithm finds a partition of the network such that the strength of edges between one core node and another core node is maximized, and the strength of edges between one periphery node and another periphery node is minimized. Diagrammatically, core-periphery structure is shown in its network formulation and in its adjacency matrix formulation in Figs. \ref{c4_fig0} A and B, respectively.

The above algorithm can be adjusted, via the incorporation of a resolution parameter $\gamma$, to select for cores of specific sizes. Here, we are interested in structural context network cores that overlap with the joint DMS-FPS network, as we motivate and explain in greater detail in Secs. \ref{inf_idx} and \ref{results2}. Given that the DMS-FPS network we use contains 140 nodes, we sought to identify cores of the same approximate size (140 nodes). Through an iterative analysis, we found that we could achieve an average core size of 140 nodes for $\gamma = 0.00045$. We implemented this resolution parameter to calculate $B'$ following Refs. \cite{Rubinov2015,Rubinov2010}:
\begin{equation}
B' = \frac{(b+b^{\textup{T}})}{2\sum_{i,j}^N B_{i,j}},
\end{equation}
where
\begin{equation}
b = B-\gamma \frac{\sum_{i,j}^N B_{i,j}}{N^2},
\end{equation}
and where \emph{N} = 400 is the total number of nodes in the network.

\subsubsection{Calculation of the influence index}
\label{inf_idx}

%\DB{Can we start this paragraph with the question we were hoping to answer or the hypothesis we wished to test, before describing the method we used to do so?} 
 
 We hypothesized that a correspondence may exist between the core of structural context networks and the subsequent rTMS induced changes in functional connections within the brain. Specifically, we hypothesized that stimulation of a region whose structural context network core coincided with the DMS and FPS networks might lead to functional change within the DMS and FPS networks. Conversely, stimulation of a region whose structural context network core did not coincide with the DMS and FPS networks might lead to diminished or no functional change within the DMS and FPS networks. To quantify the degree to which the core of a structural context network coincides with the DMS and FPS networks, we calculate the influence index (Fig. \ref{c4_fig0} C). Specifically, we compare two binary partitions where the first partition indicates core membership and the second partition indicates membership to either the FPS \emph{or} the DMS. The influence index, $\mathcal{I}$, is equal to the inverse of the sum of the absolute difference between the two binary vectors:
 
\begin{equation}
\mathcal{I} = \left( \sum_{i=1}^N \left| X_i - Y_i \right| \right)^{-1},
\end{equation}
where $X \in \mathcal{N}^{N \times 1}$ is a binary vector indicating membership to the joint FPS and DMS network, and $Y \in \mathcal{N}^{N \times 1}$ is a binary vector indicating core membership within the joint system.

\begin{comment}
The influence index is equal to the $z$-score of the Rand coefficient between the two partitions \cite{Traud2011}:
\begin{equation}
z_R=\frac{w-\mu_w}{\sigma_w},
\end{equation}
where $w$ is the Rand coefficient, $\mu_w$ is the mean of the Rand coefficient, and $\sigma_w$ is the standard deviation of the Rand coefficient. More explicitly, this statistic is calculated in the following way:
\begin{equation}
z_{R}=\frac{1}{\sigma_w}\left ( w-\frac{M_1M_2}{M} \right ),
\end{equation}
where 
\begin{multline*}
\sigma^{2}_{w}=\frac{M}{16}-\frac{(4M_1-2M)^2(4M_2-2M)^2}{256M^2}+\frac{C_1C_2}{16n(n-1)(n-2)} \\
+\frac{\left [ (4M_1-2M)^2-4C_1-4M \right ]\left [ (4M_2-2M)^2-4C_2-4M \right ]}{64n(n-1)(n-2)(n-3)},
\end{multline*}
where
\begin{equation*}
C_1=n(n^2-3n-2)-8(n+1)M_1+4\sum_{i}n^3_{i\cdot},
\end{equation*}
\begin{equation*}
C_2=n(n^2-3n-2)-8(n+1)M_2+4\sum_{j}n^3_{\cdot j}.
\end{equation*}
Here, $n_{ij}$ is the number of nodes assigned to group $i$ in partition 1 and assigned to group $j$ in partition 2, $n_{i\cdot }=\sum_j n_{ij}$, $n_{\cdot j}=\sum_i n_{ij}$, $M_1=\sum_{i}\binom{n_{i\cdot }}{2}$, $M_2=\sum_{j}\binom{n_{\cdot j}}{2}$, $M$ is the total possible number of nodal pairs, $n$ is the total number of nodes, and $w=\sum_{ij}\binom{n_{ij}}{2}$. A MATLAB implementation of this statistic is publicly available (http://netwiki.amath.unc.edu). %\DB{I think C in the above equatinos has not been defined.}
\end{comment}

\subsection{Transcranial magnetic stimulation}

\subsubsection{TMS equipment}
A primary goal of this work is to investigate how rTMS alters functional connectivity, and the role of the stimulated region's white matter connections in this process. To achieve this goal, we used repetitive TMS (rTMS) to induce changes in localized brain activity. We delivered rTMS with a Magpro X100 stimulator (Magventure; Farum, Denmark) which we connected to an air cooled MRI compatible Magventure MRI-B91 TMS coil. To map a subject's MRI image to locations on the subject's scalp, we used a stereotaxic system for neuronavigation (Brainsight; Rogue Research, Montreal, Quebec, Canada). We marked the stimulation sites on a lycra swim cap, locating the sites with a Polaris optical position Vicra camera. The location for each site was determined by subject-specific functional connectivity values, as we describe in more detail below. rTMS pulse delivery and MRI scan pulse timing were provided by a dedicated windows PC through E-prime 2.0 (Psychology Software Tools, Sharpsburg Pennsylvania USA).

\subsubsection{TMS targeting}
Regions selected for stimulation were chosen to replicate a prior study that examined how stimulation to the cortex can modulate activity in deeper regions implicated in affective disorders \cite{Oathes2018}. Here, we will provide a brief summary of the targeting strategy as it will allow the reader to better understand our use of the collected data \cite{Oathes2018}. Regions for stimulation were based on seed-based functional connectivity with either (1) the subgenual anterior cingulate cortex (sgACC) or (2) the basolateral amygdala. These two regions were initially selected due to their hypothesized importance in affective disorders. Due to individual differences in functional connectivity, this experimental protocol resulted in each subject being stimulated at different cortical locations.
%Dysfunction in the amygdala has been related to social phobia and depression \cite{Etkin2007,Hamilton2012}, and sgACC  is associated with clinical depression \cite{Greicius2007}. To narrow focus only to strong functional connections, sites with weak functional connectivity defined as absolute value $z$-scored functional connectivity strengths smaller than 0.25, were not considered \cite{Oathes2018}.
%After applying this threshold, 11 subjects were targeted for the sgACC, and 21 subjects were targeted for the amygdala. \DB{I am not following how the functional connectivity was used to determine which people got sgACC stim and which got amygdala stim. Can we unpack the choice process more?} The TMS coil was placed for a posterior to anterior induced current \DB{Why? Add ref to justify.}. 
Each subject's motor threshold was established by observing the movement of either (1) the first dorsal interosseuous muscle or (2) the abductor pollicis brevis muscle in the dominant (right) hand. The motor threshold is the minimum necessary rTMS dosage to elicit a visible motor response. This threshold is a robust and reliable response that is used to determine the intensity of rTMS dosage to other areas \cite{westin2014determination}. The threshold was set when either muscle responded by twitch to 5 of 10 subsequent rTMS stimulations, and experimental rTMS was subsequently applied at 70\% motor threshold \cite{Oathes2018}. Dosage levels up to 120\% of resting motor threshold have been found safe for clinical use \cite{hadley2011safety}.

%Region masks were obtained through FreeSurfer: each subject's high resolution T1 was processed using FreeSurfer's `recon-all' pipeline. The amygdala ROI was obtained from the Harvard-Oxford subcortical atlas and the sgACC ROI was the subcallosal gyrus ROI from the Destrieux atlas.

%\DB{This is the first time we hear about depression. I think we need to be clear at the top of this TMS targeting subsection that the data were taken from a previous study that was interested in depression, and that many of the choices in terms of data collection are motivated by that goal. Otherwise, a bunch of the description here seems ad hoc and unjustified.} TMS target areas were chosen by considering that area's resting state connectivity to two seed regions of interest. The sgACC seed was chosen in light of a series of studies demonstrating a link between sgACC activation and severity of major depressive disorder clinical symptomatology, as well as an association between TMS stimulation at that site and symptomatology \cite{Liston2014}. The amygdala seed was chosen in accordance with prior work \DB{Can we provide more of a motivation, justification, or reason here? I don't know what "in accordance with prior work" really means.} \cite{Etkin2009}. Region masks were obtained through FreeSurfer: each subject's high resolution T1 was run through FreeSurfer's `recon-all' pipeline. The amygdala ROI was obtained from the Harvard-Oxford subcortical atlas and the sgACC ROI was the subcallosal gyrus ROI from the Destrieux atlas.

\subsection{Experimental Protocol}
Subjects first underwent a baseline structural (T1) and fMRI resting scan to enable localization of rTMS targets. On a separate day, subjects underwent a pre-rTMS resting scan as described in more detail above. Immediately following the resting state scan, subjects received a round of intermittent theta burst (iTBS) TMS \cite{huang2005theta}, which consisted of 50 Hz burst pulses of 3 stimulations each. The sequence was repeated at 5 Hz in 20 blocks of 10 sequences, amounting to 600 pulses (1800 stimulations) total over a time period of approximately 4 minutes. Approximately 7 minutes following neuromodulation (during which interleaved rTMS/fMRI with non-neuromodulatory single TMS pulses were delivered; data reported elsewhere \cite{oathes2021resting}), subjects underwent a post-rTMS resting state scan.

\subsection{Predicting changes in functional connectivity}

\subsubsection{Model design}
A primary goal of this work is to establish whether considering structural connectivity improves predictions of changes in functional connectivity in response to rTMS. To achieve this goal, we construct and test two models that predict changes in the strength of the aggregated functional connection between the DMS and the FPS as a result of rTMS. One model includes information about structural linkages---via the structural context network---whereas the other does not. Our first model, which does not incorporate structural linkage information, is:
\begin{equation}
\Delta \textup{FC}_{ij}=f_{1} \left(\Delta \textup{Act}^{(k)} \right ),
\end{equation}
where $\Delta \textup{FC}_{ij}$ is the magnitude of change in the strength of the functional connection between large-scale areas $i$ and $j$, $f_1$ is a linear function that we define below, and $\Delta \textup{Act}^{(k)}$ is the change in activity of the region $k$ to which rTMS is applied. In accordance with prior work, activity is defined as the statistical variance of the  BOLD timeseries \cite{Murphy2019}. In this model, the large-scale area $i$ is the entire FPS, and the large-scale area $j$ is the DMS; thus, the functional connectivity between these two large-scale areas is defined as the mean weight of all edges linking a region in the FPS to a region in the DMS. The linear function $f_1$ is
\begin{equation}
f_1 \left (x \right ) = a_0 + a_1 x + \epsilon,
\end{equation}
 where $f_1$ is a linear function of $x$ with intercept $a_0$, linear coefficient $a_1$, and error $\epsilon$. With this model, we can quantify how accurately we can predict changes in functional connectivity using changes in regional activity alone.%\DB{Can we fill out the following statement: "With this model, we can test the hypothesis that blah blah blah."?}

Our second model, which incorporates structural linkage information, is:
\begin{equation}
\Delta \textup{FC}_{ij}=f_2 \left(\Delta \textup{Act}^{(k)},\textup{SC}^{(k)}_{ij} \right ), \label{eq5}
\end{equation}
where $\textup{SC}^{(k)}_{ij}$ is defined above (Eq. \ref{eq1}). Notably, this second model includes the structural context of the stimulated region $k$. The linear function $f_2$ is
\begin{equation}
f_2\left (x,y \right ) = a_2 + a_3 x + a_4 y + a_5 x y + \epsilon,
\end{equation}
where $f_2$ is a linear function in $x$, $y$, and the interaction between $x$ and $y$. With this model, we can quantify how accurately we can predict changes in functional connectivity using changes in both magnitude of regional activity and structural connectivity. 

To summarize, $f_1$ and $f_2$ are both predictive models of change in functional connectivity, where $f_2$ contains structural information whereas $f_1$ does not. Our hypothesis, which we test below, is that $f_2$ will produce more accurate predictions than $f_1$.

\subsubsection{Model testing} \label{mod_test}
To compare these two models, we calculate two statistical differences between the pre- and post-rTMS data. First, we calculate the mean connectivity between the DMS and FPS in the post-TMS scan and subtract from it the mean connectivity between the DMS and FPS in the pre-rTMS scan; we then take the absolute value of the difference. We denote this variable as $\Delta \textup{FC}_{ij}$. Second, from the same dataset and for each individual, we also calculate the difference in pre- and post-rTMS root mean square of the stimulated region $k$'s timeseries \cite{Murphy2019}, which we denote as $\Delta \textup{Act}^{(k)}$. Comparing the two models also requires us to have an estimate of each region's structural context $SC^{(k)}_{ij}$. The TMS dataset does not include diffusion imaging, and therefore we instead chose 100 randomly selected subjects from the HCP, calculated the structural context of each region for each of the 100 structural matrices, and then calculated the mean structural matrix across those 100 estimates as the representative structural context.

To compare the performance of our two models, we first fit the models $f_1$ and $f_2$ to the data described above. Next, we calculated the Bayesian information criterion (BIC) to assess performance between the two models. Notably, BIC accounts for the fact that $f_2$ contains an additional parameter relative to $f_1$, and therefore may be expected to perform at least as well as $f_1$. We fit each model to the data 100,000 times, where during each iteration we drew a sample of 29 subjects with replacement from the pool of 29 subjects. After each iteration, we calculated the BIC for each model. After completing all 100,000 iterations, we calculated the mean BIC for each model. Finally, we calculated the difference in mean BIC to assess which model had lower BIC; the lower BIC model is the preferred model.

\subsection{Calculation of correlations and significance}
Unless otherwise stated, we used the following protocol to calculate the degree and significance of correlation between two sample distributions $X \in \mathcal{N}^{m \times 1}$ (i.e., regional activity) and $Y \in \mathcal{N}^{m \times 1}$ (i.e., DMS-FPS functional connectivity). We followed a bootstrapping procedure where, for 30,000 bootstrap iterations, we randomly drew the same set of $m$ randomly drawn rows (with replacement) from $X$ and $Y$ to generate resampled vectors $X'$ and $Y'$. For each iteration $i$, we then calculated the Pearson correlation coefficient $r(i) = corr(X',Y')$ where $i = \{1,30,000\}$. From this vector, $r$, containing Pearson correlation coefficients of bootstrap resamples, we report the mean correlation coefficient, as well as the significance $p = \textup{Pr}(r<0)$.

\section{Results} \label{results1}

\subsection{Models that include structural linkages better predict strength of functional connectivity}

\begin{figure}
\centerline{\includegraphics[width=.7\textwidth]{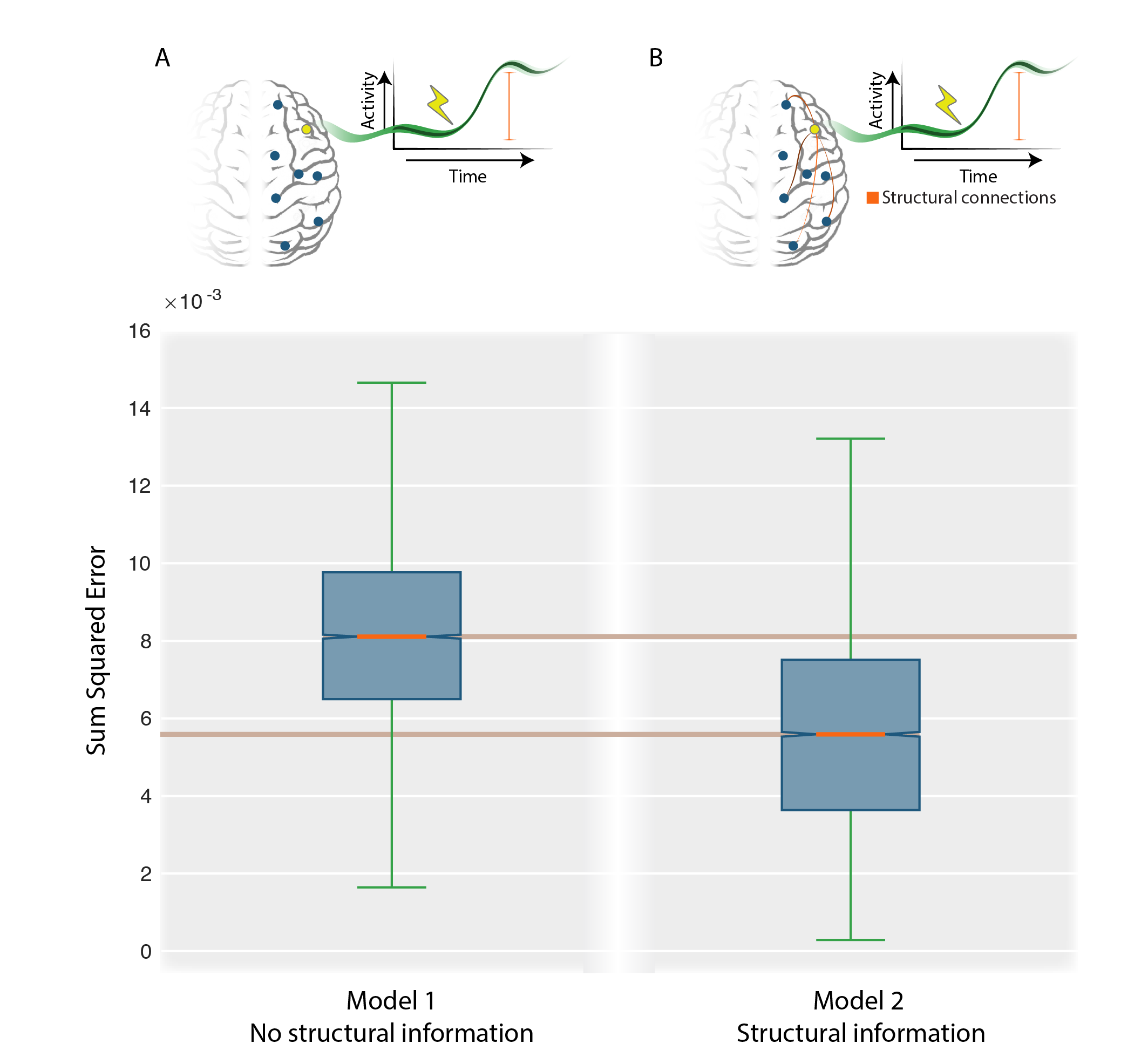}}
\caption{\textbf{The model incorporating structural linkage outperforms the model that is structure-naïve in predicting functional coupling changes induced by rTMS.} We found that the mean SSE for the structurally informed model (mean = 0.0056) was significantly smaller than the mean SSE for the structurally naïve model (mean = 0.0081) ($F = 5298, \textup{df} = 19998, p<0.001$) indicating that the predictions of model $f_2$ are more accurate than those of model $f_1$. \emph{(A)} (Top) Schematic of structure-naïve linear model. (Bottom) Sum of squared error distribution for structure-naïve linear model. \emph{(B)} (Top) Schematic of structure-informed linear model. (Bottom) Sum of squared error distribution for structure-informed linear model. The central mark on the boxplots indicates the median value, and the box extends to the 25th and 75th percentiles. The notches and horizontal dark orange lines indicate the 95th percent confidence interval of the medians.%\emph{(A)} We fit model $f_1$, which did not include structural information, to the data using 10,000 bootstrap resamples to generate a distribution of \emph{SSE} (mean $SSE = 0.0159$, standard deviation = 0.0035). \emph{(B)} We fit model $f_2$, which did incorporate structural information, to the data using 100,000 bootstrap resamples to generate a distribution of \emph{SSE} model fits (mean $SSE = 0.0102$, standard deviation = 0.0029). We found that, for a given bootstrap sample, the probability that $f_1$ produces a smaller SSE than $f_2$ is less than 0.001.
}
 \label{c4_fig1}
\end{figure}

Here we hypothesize that a linear model informed by TMS-induced regional activity changes will predict changes in functional connection strength. Further, we hypothesize that the addition of structural white matter information to the model will improve model performance. To test and compare the performance of our models, we used the procedure described in Sec. \ref{mod_test}. 
%We found that $\textup{Pr}(\textup{SSE}_{f_2}>\textup{SSE}_{f_1}) < 0.001$, 
We found that the mean sum squared error (SSE) for the structurally informed model (mean = 0.0056) was significantly smaller than the mean SSE for the structurally naïve model (mean = 0.0081) ($F = 5298, \textup{df} = 19998, p<0.001$) indicating that the predictions of model $f_2$ are more accurate than those of model $f_1$ (Fig. \ref{c4_fig1}). It is important to note that model $f_2$ includes more parameters than model $f_1$. The fact that model $f_2$ outperforms model $f_1$ could be due trivially to this fact or could be due to greater sensitivity to the actual biology \cite{theil1961economic}. To determine which of these possible explanations was true, we implemented a complementary approach to determine whether the addition of structural linkage information improves the prediction of functional connectivity changes by calculating model BIC (see Methods). We found that the mean difference in BIC between the two models was 5.5933, indicating a preference for the model containing structural information.

%Specifically, we created a non-parametric permutation-based null dataset by shuffling structural context values uniformly at random across subjects. While preserving the number of model parameters, this process serves to break the correspondence between (i) structural linkages of the stimulated region and (ii) both the change in functional connection strength and the change in activity. As described in Sec. \ref{mod_test}, we fit model $f_2$ to this new null dataset and generated a new set of SSE values, $SSE'_{f_2}$. We found that $\textup{Pr}(\textup{SSE}_{f_2}>\textup{SSE}'_{f_2}) < 0.05$, indicating that the predictions of model $f_2$ are more accurate when fit to the initial dataset \emph{versus} the null dataset.

%Model $f_2$ has a higher mean $r^2$ \DB{r-squared of what with what? Need to clearly state the two variables here.} than model $f_1$ (0.717 \emph{versus} 0.141), indicating that model $f_2$ explains more variance in DMS-FPS functional connectivity than model $f_1$. To more directly compare these models, on each bootstrap resample we also calculate the difference in $r^2$ to produce a distribution of differences in fit ($r^2(f_2) - r^2(f_1)$ ) (Fig. \ref{c4_fig1} C). We find that model $f_2$ significantly outperforms model $f_1$ in terms of $r^2$ value ($p < 0.0001$). \DB{How was the p-value calculated? The p-value is associated with the r-squred of f2? Or the p-value is associated with the differences in r-squared between f2 and f1? Unclear.}

%We find that the model fits the true data (Fig. \ref{c4_fig2} A) more accurately than it fits the null data (Figs. \ref{c4_fig2} B and C). \DB{Clarify again how we are measuring model fit? r-squared between what and what precisely?} 

Note that here we focused our attention on the change in functional coupling between the DMS and FPS. We found that the way in which a stimulated region is structurally coupled to those systems influences the resulting change in functional connection strength between these systems in response to rTMS. This finding indicates that there may be a correspondence between the DMS-FPS structural coupling of the stimulated region and the DMS-FPS functional coupling.

%\begin{figure}
%\centerline{\includegraphics[]{fig2.png}}
%\caption{\textbf{Model $f_2$ outperforms model $f_2$ fit to permuted data.}  \DB{Title would probably be more accessible if we used words, as I mentioned in the caption for Figure 1.} \emph{(A)} We fit model $f_2$, which incorporates structural information, to the data using 30000 bootstrap resamples to generate a distribution of $r^2$ model fits (mean $r^2 = 0.71676$ \DB{add standard deviation}). \emph{(B)} In evaluating a non-parametric permutation-based null model, we again fit model $f_2$ to the data, but only after permuting structural coupling data uniformly at random across subjects, using 30000 bootstrap resamples to generate a distribution of $r^2$ model fits (mean $r^2 = 0.29758$ \DB{add standard deviation}). \emph{(C)} This plot shows the difference in $r_2$ for each bootstrap resample: specifically, model $f_2~r^2 - $ permuted model $f_2~r^2$. We observed that model $f_2$ applied to the real data has a significantly higher $r^2$ than the same model applied to the permuted data ($p = 0.026$). \DB{explain how the p-value was obtained in the new Methods subsection where you will describe the bootstrapping procedure.}}
 %\label{c4_fig2}
%\end{figure}

\subsection{Topology of structural context network predicts TMS-induced changes in functional connectivity.} \label{results2}

To assess the generalizability of our findings, we broadened our investigation from large-scale functional coupling between systems to smaller scale functional coupling between regions. For each of the 29 individual TMS targets $k$, we used Eq. \ref{sc} to calculate the structural context of that region $k$ with respect to each pair ($i$, $j$) of the 400 cortical regions in the Schaefer atlas. Through this process, we created a structural context \emph{matrix} or \emph{network}, in contrast to the single value obtained in the previous section.
 %(Fig. \ref{c4_fig3} A). 
 Likewise, we also calculated the pairwise change in functional connectivity from the pre-rTMS resting state scan to the post-rTMS resting state scan ($\Delta \textup{FC}$) to quantify the degree of functional change that accompanied the regional TMS stimulation.% (Fig. \ref{c4_fig3} C).

%Using these estimates, we first sought to determine to what degree the $\Delta \textup{FC}$ network induced by stimulating region $k$ reflects region $k$'s structural context network. To quantify this relationship, we calculate the Spearman correlation between the functional change network and structural context network.
%two networks: for example, the correlation between the upper triangle of the matrix in Fig. \ref{c4_fig3} A and the upper triangle of the matrix in Fig. \ref{c4_fig3} C). 
%We refer to this correlation as the \emph{matching} between the two networks. %Interestingly, matching values varied markedly%(Fig. \ref{c4_fig3} D)
%, with some being positive and some being negative.
 Next, we sought to determine whether there were certain characteristics of the structural context network that explained the associated functional changes. Prior work in models of communication in brain networks \cite{Avena-Koenigsberger2018a} suggests that the existence of many long distance paths---such as those found in rich-clubs and more general core-periphery structures---can support the broad propagation of perturbative signals. Accordingly, we hypothesized that structural context networks with strong core-periphery structure would be associated with more pronounced functional changes. %We find that many of the structural context networks exhibit strong core-periphery structure. 
 Consistent with our hypothesis, we find that the core-ness (Eq. \ref{eq_coreness}) is positively correlated with total cortical functional change ($r = 0.37$, $p = 0.007$). Here, total cortical functional change is the mean change in functional connectivity of all pairwise connections. Stated another way, for networks with pronounced core-periphery structure, acute rTMS-induced functional network changes may be partially predicted by observing structural network architecture alone. These results suggest that the location of the stimulated region within the broader structural network directly impacts its potential to induce wide-spread changes in functional connectivity.

\begin{figure}
\centerline{\includegraphics[]{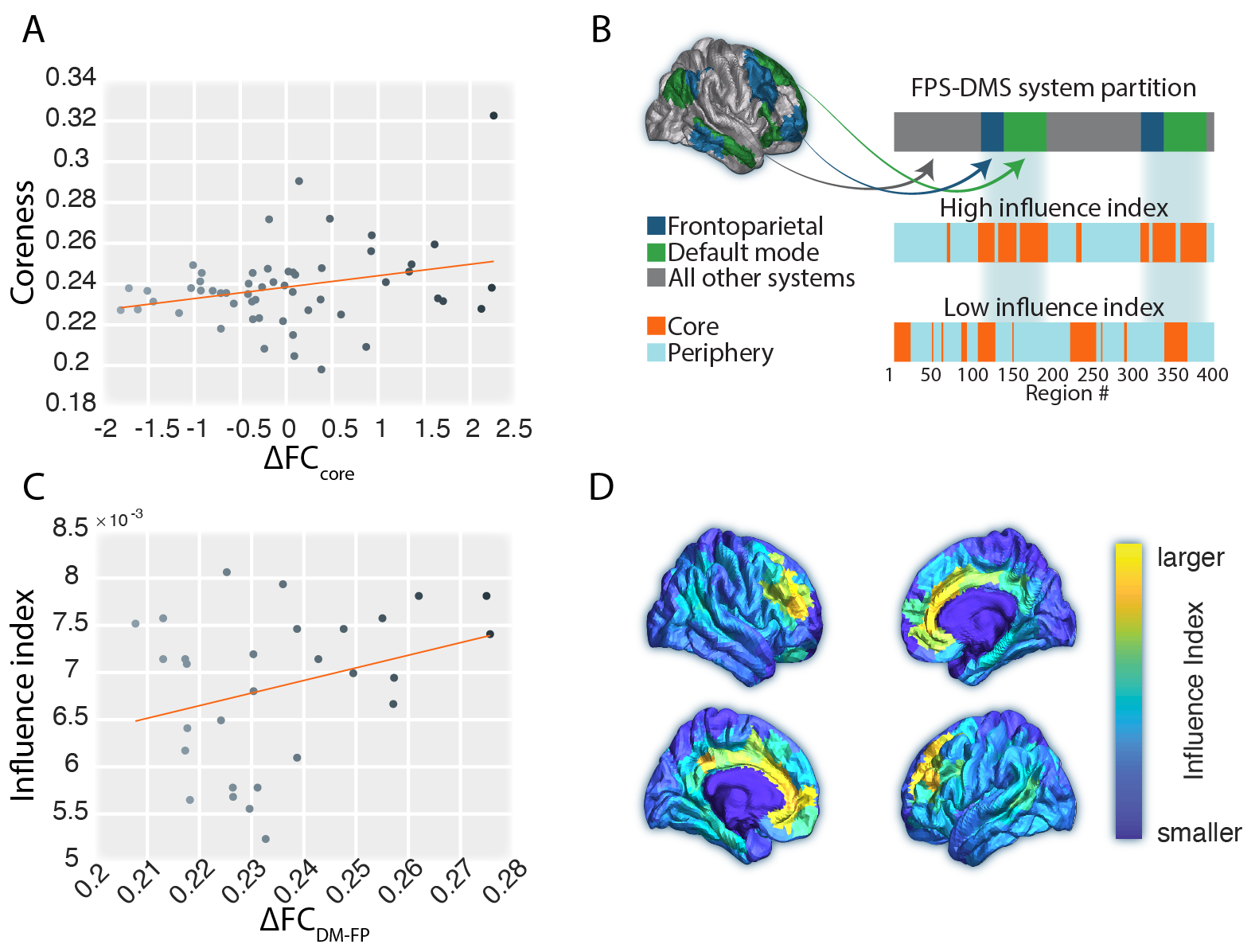}}
\caption{\textbf{Core-periphery architecture predicts functional changes.}  \emph{(A)} The coreness of a structural context network is related to the magnitude of functional change in the core following acute rTMS (\emph{r} = 0.191, \emph{p} = 0.006). \emph{(B)} The joint frontoparietal and default mode partition represented on the brain surface and as a vector. This joint partition is composed of all non-gray regions. The \emph{influence index} quantifies the similarity between the system partition and a core-periphery partition. A core-periphery partition that overlaps more with the system partition has a high influence index, whereas a core-periphery partition that overlaps less with the system partition has a low influence index. \emph{(C)} The influence index of a stimulated region is related to the mean absolute change in the weight of edges linking a region in the DMS to a region in the FPS (\emph{r} = 0.378, \emph{p} = 0.0031). \emph{(D)} The influence index of all 400 brain regions projected onto the cortical surface.}
\label{c4_fig4}
\end{figure}

Motivated by prior work that has underscored the relevance of absolute changes in functional connectivity \cite{Fox2012c,Lynall2013,Konrad2010}, we next sought to characterize more precisely how core-ness was related to change in functional connectivity. We approached this question by first calculating the core-periphery partition of the structural context network for each subject. % (Fig. \ref{c4_fig3} B).
Then we performed two analyses. First, we investigated the relationship between core-ness and the absolute mean change in strength of edges linking a core node to another core node. Second, we investigated the relationship between core-ness and the change in strength of edges linking a periphery node to another periphery node. We found no correlation between core-ness and the absolute change in functional connectivity within the periphery ($r = 0.1364$, $p  = 0.0612$). In contrast, we did find a significant positive correlation between core-ness and the absolute change in functional connectivity within the core (Fig. \ref{c4_fig4} A; $r = 0.191, p = 0.006$). Further, for each of the 29 stimulated regions we calculated the structural context network such that each stimulated region \emph{x} had a corresponding structural context network $SC_x$. For all 29 regions, we found that the region \emph{x} was included in the core of $SC_x$. This indicated that stimulation was always applied within the core of the structural context network. These results further refine the relationship between the structural connections of the stimulated region and downstream functional connectivity changes induced by acute rTMS, and suggest a principled way to influence a specific set of functional connections: stimulate a region where the core corresponds to the desired set of functional connections for a particular system. The potential practical applications of this finding are discussed in Sec. \ref{discussion1}.

%\DB{Need some big-picture conclusion statement similar to the one I put at the end of the previous paragraph.} 

%\DB{The following two sentences seem to belong in the Methods section, rather than here in the Results. Can you move them to an appropriate spot? And if you need to add more contextualizing sentences around them, feel free to do so there.} We calculated a structural context network for each region $k$, and found regions that belonged to the core of that structural context network. The region $k$ was always included in the set of core regions of region $k$'s structural context network.

%\begin{figure}
%\centerline{\includegraphics[]{fig3.png}}
%\caption{\textbf{Relationship between structural linkages and functional connectivity.}}  \emph{(A)}  An example structural context network for region \emph{k}. \emph{(B)} The same structural context network reordered to group core nodes together and periphery nodes together. \emph{(C)} A corresponding network displaying change in functional connectivity pre- \emph{versus} post-TMS, where TMS was delivered at region \emph{k}. \emph{(D)} A histogram showing the observed matching values from stimulation at all 32 individual stimulation points. \emph{(E)} The matching metric is positively correlated with core-ness ($r = 0.29$, $p = 0.361$ \DB{this p-value is not significant. Is there an issue here?}). \emph{(F)} Absolute average change in the weight of functional edges linking a core node to another core node is positively correlated with the core-ness ($r = 0.3188, p = 0.0392$).
%\label{c4_fig3}
%\end{figure}

\subsection{Core-periphery structure predicts changes in default mode - frontoparietal coupling} \label{results3}

Results from the prior section suggest that stronger core-periphery organization in a region's structural context network is related to greater change in functional coupling in core nodes. We used these findings at the regional level to better understand changes in functional coupling at the systems level. Returning to our earlier investigation of the functional connection between the DMS and the FPS, we hypothesized that the greatest changes in the DMS-FPS functional connection would occur when the core of the stimulated region is maximally overlapping with the DMS and FPS. To test this hypothesis, we calculated the core-periphery partition for each stimulated region $k$'s structural context network. We then calculated the \emph{influence index} of region $k$ by quantifying the degree of overlap between the core and the FPS and DMN (Fig. \ref{c4_fig4} B; see Methods). Finally, we quantified the mean absolute change in the weight of edges linking a region in the DMS to a region in the FPS. 

When considering the edges linking a region in the DMS to a region in the FPS, we found that change in edge weight was positively related to the influence index (Fig. \ref{c4_fig4} C; $r = 0.378$, $p = 0.0031$).  Building on results from the prior section, these findings further support a link between the structural connections of the stimulated brain region and downstream functional connectivity changes in response to neuromodulation. Specifically, the more similar the core of the stimulated region's structural context network is to the set of functional connections to be stimulated, the larger the magnitude of change in that set of functional connections. Given that result, we can now ask the question: which regions of the brain have the highest influence index? In other words, if we would like to alter the FPS-DMS functional connections, where might we stimulate to maximally change this set of functional connections? To answer this question we calculated the influence index for each of the 400 parcel regions (Fig. \ref{c4_fig4} D). We found that the influence index is highest bilaterally in the dorsolateral prefrontal cortex, medial and dorsomedial prefrontal cortex, and anterior cingulate cortex. In summary, the above findings suggest that the topology of the stimulated region's structural context network partially predicts which functional connections might be most altered.

%medial precuneus.

%Finally, we quantify the mean absolute change in the weight of (i) edges linking two regions within the DMS, (ii) edges linking a region in the DMS to a region in the FPS, and (iii) edges linking two regions within the FPS. 

%When considering all three groups of edges together, we found that the influence index was significantly positively correlated with the magnitude of the functional change (Fig. \ref{c4_fig4} B, $r = 0.415$, $p = 0.0028$). When considering just the edges linking a region in the DMS to a region in the FPS, we found that change in edge weight is positively related to the influence index ($r = 0.299$, $p = 0.021$). We observed no significant relationship between the influence index and the change in the weight of edges linking two regions within the DMS ($r = 0.058$, $p = 0.362$), or change in the weight of edges linking two regions within the FPS ($r = 0.3094$, $p = 0.0634$). \DB{Can we add a sentence here to motivate the analysis and results your described in the next two sentences?} Lastly, we calculate the influence index for each of the 400 parcel regions to quantify the degree to which the core of each region's structural context network overlaps with the FPS - DMS partition (Fig. \ref{c4_fig4} C). We found that the overlap is most notable bilaterally in the dorsolateral prefrontal cortex, medial and dorsomedial prefrontal cortex, and medial precuneus.

\section{Discussion}
%\DB{Add a big picture motivation and brief summary of findings here. One paragraph.}
Functional connection strengths derived from resting-state fMRI partially explain individual variations in symptom severity across a number of neuropsychiatric disorders \cite{korgaonkar2020intrinsic,Fox2012c,Lynall2013,Konrad2010}. Directly altering FC via noninvasive neuromodulation as a means to alter symptoms is a promising therapeutic approach \cite{Fox2012d,Liston2014,sliwinska2020dual,goldapple2004modulation}. However, methods to select a cortical stimulation point to achieve a given FC alteration are underdeveloped. Here, we suggest one possible approach informed by TMS-induced regional activity change coupled with white matter tract architecture. By implementing our approach, one can estimate the expected change in strength due to TMS of a given FC. We found that we could more accurately estimate TMS-induced FC changes when our approach was informed by regional activity changes \emph{and} white matter architecture \emph{versus} regional activity change alone. Further analysis suggested that predictive improvement attributed to white matter architecture may be partially explained by the core-periphery organization of the white matter networks. Specifically, we found that stimulating a region whose structural context network had strong core-periphery organization resulted in more pronounced FC changes in the core, but not in the periphery. Finally, we generalized this result to demonstrate that stimulating a region where the core of its structural context network coincided with the DMS and FPS resulted in more pronounced FC changes between the DMS and FPS. Given that the DMS-FPS FC strength is correlated with individual differences in severity in a number of clinical symptoms, including deficits in working memory \cite{Murphy2019}, these findings suggest a principled approach to selecting a stimulation point to alter those symptoms: stimulate at a region whose influence index for the DMS-FPS is high.

\subsection{Models that include structural linkages better predict strength of functional connectivity.}

Precisely how regional evoked activity relates to evoked functional connectivity changes remains an open question in human neuroimaging. Efforts to address this question have been motivated most recently by two simultaneous advances: (1) the development of network neuroscience, which has emphasized the utility of interregional functional connections as a correlate of neuropsychiatric symptoms, and (2) the widespread adoption of TMS as a tool to alter either regional activity or interregional functional connections \cite{oathes2021combining}. Alterations in functional connectivity are associated with a wide variety of neuropsychiatric symptoms, including those of Alzheimer's disease \cite{schumacher2019dynamic,Fox2012c}, schizophrenia \cite{Lynall2013}, and ADHD \cite{Konrad2010}. Likewise, interest is rapidly growing in the question of how to use TMS to alter functional connectivity \cite{Fox2012c}. TMS has been used to alter the functional architecture of the motor system \cite{Grefkes2010}, the functional connection from the dorsolateral prefrontal cortex (DLPFC) to the cingulate \cite{Paus2000}, and the functional connection from the sensory cortex to the motor cortex \cite{Pleger2006}.

Notably, these previous studies assess the associations between changes in regional activity and changes in functional connectivity, but do not account for the white matter tracts that physically link brain areas with one another. Yet, a separate basic science literature has clearly demonstrated the degree to which white matter shapes functional connectivity: functional network architecture, during both task \cite{baum2020development,Hermundstad2013} and rest \cite{Skudlarski2008,Greicius2009}, displays remarkable similarities to the underlying structural network architecture. These results suggest that white matter architecture may be useful in explaining properties of functional networks \cite{Honey2010}, which in turn suggests that structural networks may shape responses to neuromodulatory interventions \cite{korgaonkar2021white}, and may prove useful in designing those interventions \cite{sydnor2022cortical}.

To determine whether considering structural networks improves the prediction of functional changes following stimulation, we compared the accuracy with which two different models predicted changes in the DMS-FPS functional connection in response to neuromodulatory rTMS. Agreeing with out expectations, we found that the model which incorporated structural linkage information was more accurate than the model which did not incorporate this information. Our findings complement prior evidence that white matter explains \emph{variance} in functional connection strengths, and further extend the state of knowledge by demonstrating that white matter predicts \emph{changes} in functional connection strengths. Our study is therefore a first step toward the goal of accurately predicting downstream changes in functional connectivity resulting from therapeutic TMS, by demonstrating the utility of structural networks in predictive models.

\subsection{Core-periphery organization of structural context networks predicts functional connectivity changes.}
\label{discussion1}

Core-periphery networks are composed of a central group of densely connected nodes (the core) surrounded by a second group of nodes (the periphery) that connect to the central group but not to each other. Core-periphery structure has been previously observed in brain network dynamics \cite{Bassett2013a}, social networks \cite{Doreian1985}, and the social networks of research scientists \cite{Brieger1975}, among other networked systems \cite{Rombach2014}. Core-periphery organization may impart robust adaptability to the network \cite{Kitano2004} by serving two functional roles \cite{Bassett2013a}: the core may provide functional stability, whereas the periphery may provide functional flexibility to changing environmental demands \cite{Otokura2016,Ekman2012}.

\begin{figure}[H]
%\centerline{\includegraphics[]{fig0-01.png}}
\centerline{\includegraphics[width=.8\textwidth]{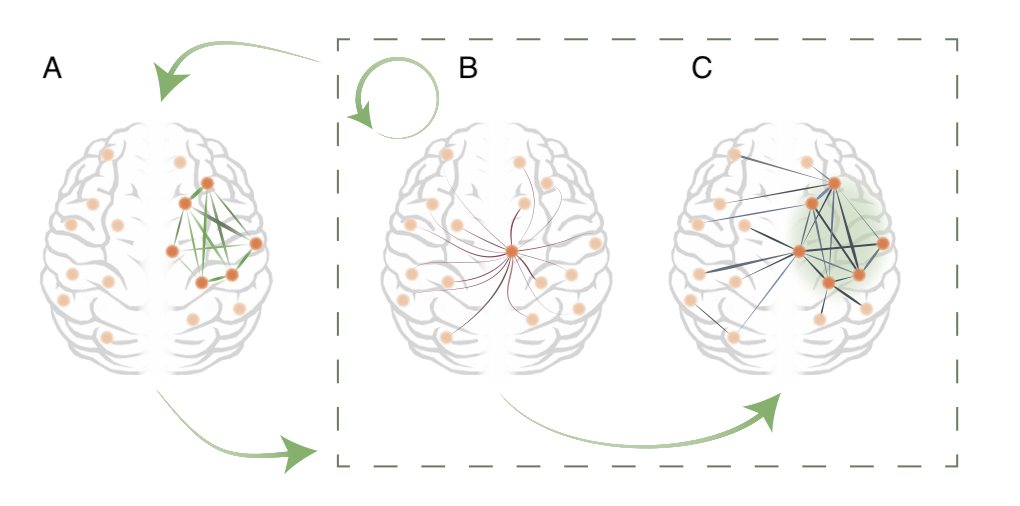}}
\caption{\textbf{Procedure to select a stimulation region to modulate a set of functional connections.} \emph{(A)} First, the desired set of connections to be modified are identified. Next, the procedure within the box is repeated by iteratively visiting each potential stimulation point \emph{(B)} to determine the core of that region's structural context network \emph{(C)}. The region whose structural context core best aligns with the desired set of functional connections to alter would be the optimal region to stimulate.} 
 \label{iteration}
\end{figure}

%\DB{Need one more sentence of intro to motivate the result you are about to report. Maybe something like "To test this possibility, we X. We found Y."?} 
Motivated by this putative functional role of the core, we tested the hypothesis that stimulating networks with more densely (structurally) connected cores may result in greater functional change within those cores. We found that stronger core-periphery organization (core-ness) in the structural context network correlated with the amount of change in functional connection strength in the core following rTMS. Under several models of dynamics, the core is an area where activity traversing the network may linger because connections are predominantly located among core nodes rather than among periphery nodes. While this feature has utility in detection algorithms \cite{Cucuringu2016}, it also has important consequences for the propagation of activity induced by stimulation: such activity may have a localized, strong, and stable effect in the core, and a diffuse, weak, and transient effect in the periphery. Our results suggest that to alter a set of functional connections, one could stimulate a region where the structural context core overlaps with those functional connections. Practically, this approach could be implemented by a procedure in which the desired connections to be modified are first identified (Fig. \ref{iteration} A), and then by iteratively visiting each potential stimulation point (Fig. \ref{iteration} B) to determine the core of that region's structural context network (Fig. \ref{iteration} C). According to the above results, the region whose structural context core best aligns with the desired set of functional connections to alter would be the optimal region to stimulate.

\subsection{Core-periphery structure predicts changes in coupling between the default mode and frontoparietal systems.}

%\DB{Start with motivation before reporting a result. Need one more sentence of intro to this paragraph.} 
%We defined a joint network containing functional connections (1) from a region in the DMS to another region in the DMS, (2) from a region in the DMS to a region in the FPS, and (3) from a region in the FPS to another region in the FPS. 

To directly test the above suggestion, we calculated the \emph{influence index} to quantify the degree of overlap between the core and the DMS and FPS networks. We hypothesized that the greater the influence index, the greater the change in the functional connection between the DMS and FPS. We found that the influence index was correlated with the amount of change in functional connections between the DMS and FPS networks following rTMS. Our finding suggests a particular relationship between structural connectivity and changes in functional connectivity that could inform strategies to modulate functional connections via brain stimulation. Specifically, to modulate the set of connections listed above, one could consider stimulating the cortical region for which the influence index is largest. Notably, here we chose to select cortical regions that are accessible by TMS \cite{Oathes2018,valchev2015ctbs}, but deeper structures may be targeted using other techniques such as deep brain stimulation \cite{Fridgeirsson2020,Mueller2018} or focused ultrasound \cite{Sanguinetti2020}. %\DB{Add some refs to this paragraph.}

%\DB{need opening sentence motivating the influence index.}
After establishing a relationship between the influence index and change in FC, we sought to generalize these findings by determining the influence index over the entire cortex. In this way, we obtained a broad sense of which cortical regions, when stimulated, maximally alter FC. We found three broad areas that tended to have a high influence index over the strength of the functional connection between the DMS and FPS: (1) bilateral DLPFC, (2) bilateral medial and dorsomedial prefrontal cortex (DMPFC), and (3) bilateral anterior cingulate cortex (ACC). These results suggest that TMS applied to any of these three regions may alter the functional connection between the FPS and DMS more than stimulation to other cortical regions. We will discuss each in turn. 

First, supporting our finding of a high influence index in the DLPFC, activity in this region is thought to modulate the strength of the functional connection between the DMS and FPS \cite{Fox2005,Fransson2005}. Although the mechanism by which the DLPFC modulates the functional connection is not yet known, recent evidence suggests that changes in DLPFC activity do in fact \emph{cause} changes in the strength of this functional connection \cite{Chen2013}. Briefly, in an experiment by Chen and colleagues, single-pulse excitatory TMS was delivered to a region within the FPS which induced negative coupling (psychophysiological interaction) between the DMS and FPS \cite{Chen2013}, indicating a causal relationship between the two systems. Our finding of a large DLPFC influence index suggests that this causal relationship may be partially underpinned by DLPFC white matter tracts. Specifically, the architecture of DLPFC white matter fibers may structurally constrain stimulation-induced activity to remain within the FPS and DMS, concentrating effects to these systems.

Second, we found large influence index values within the ACC and MPFC. The MPFC \cite{Gusnard2001} is a central components of the DMS, while the DLPFC is a component of the FPS. Lastly, the ACC has strong connections to FPS and is important in modulating FPS functional connections \cite{di2020anterior}. The FPS and DMS are competitive systems: when activity increases in one, it decreases in the other \cite{Boveroux2010,Hannawi2015,Murphy2019}. This configuration is thought to enable swift and efficient toggling between brain states subserved by each of the two systems \cite{Dixon2017}. Our finding of a high influence index in both the DLPFC \emph{and} regions of the DMS hints at a system of mutual governance by which increases in activity within the DLPFC \emph{versus} the DMPFC or ACC may have opposite effects on the strength of the functional connection between the DMS and FPS. This configuration may allow for modulation of that functional connection, the strength of which flexibly alters in response to environmental demands \cite{Cole2013}. Our results suggest that this fast and flexible reconfiguration of the functional connection strength between the DMS and FPS may be underpinned by a specific distribution of white matter fibers, enabling two discrete brain regions to potentially have opposing effects on this functional connection \cite{Murphy2019}.% \DB{Maybe cite your Nat Commun paper in this paragraph?}

\subsection{Methodological Considerations}
Several methodological considerations are pertinent to this work. Most notably, while our findings rely on coupling fMRI and diffusion imaging, data from these two distinct modes was drawn from separate subject populations. This was a necessary concession given that the TMS experiment imaging protocol did not include diffusion imaging. Our solution to this limitation, as described in Sec. \ref{data_acquisition}, was to calculate an averaged structural network from the Human Connectome Project dataset. Therefore, we assigned the same averaged diffusion structural networks to each of the subject-specific functional data. We expect that this approach would lead to an underestimation of the effect of white matter structure on functional changes because it is insensitive to individual differences in white matter. As such, we anticipate that our results will hold for studies where diffusion data is available for each subject in the TMS experiments. Nonetheless, due to this limitation, our results present only a partial picture of the relationship between white matter, functional connectivity, and regional activations in characterizing rTMS effects.

%\DB{Need to add a full Methodological Considerations and Limitations section here.}

\subsection{Conclusion}

It is hypothesized that changes in coordinated brain activity underlie clinical improvements to treatments across neuropsychiatry. Predicting changes in functional connection strength that arise from localized brain stimulation thus has important implications in clinical neuropsychiatry. Accurate prediction would allow for better mechanistic explanations of how specific brain network changes with treatment lead to improvements of specific symptoms in specific patients. While efforts to develop predictive tools are underway, it is uncommon for these efforts to consider how structural linkages constrain the dispersion of activity across the brain. Importantly, developments in neuroimaging suggest a tight correspondence between white matter networks and functional brain networks during both task and rest. Supporting these findings, we demonstrate that the accuracy of a model predicting changes in functional connectivity in response to localized TMS is significantly improved by the addition of structural linkage information. Providing more evidence for the utility of structural linkages in predicting functional network changes, we show that the more closely the core of the structural context network overlaps with the target set of functional connections, the larger the change in those functional connections. Lastly, we demonstrated how this finding could be utilized to select a stimulation region in order to alter the functional connection strength between the DMS and FPS. In sum, this work adds to a growing body of literature which suggests a dependency of functional connectivity on structural connectivity, and discusses how this dependency may be used to alter functional connections within the brain.

\section*{Acknowledgments}

This work was largely supported by the National Institute of Mental Health R01 MH111886 (DJO), 1-RF1-MH-116920-01 (DJO, DSB, TDS), R-01-MH-113550 (TDS, DSB), R-01-EB-022573 (TDS), R-37-MH-125829 (TDS), R-01-MH-112847 (TDS). DSB would also like to acknowledge support from the  Paul  G.  Allen  Foundation,  the  Army  Research  Laboratory  (W911NF-10-2-0022),  the  Army  Research  Office  (Grafton-W911NF-16-1-0474), and the National Science Foundation (PHY-1554488,  BCS-1631550, and IIS-1926829).  The  content is solely the responsibility of the authors and does not necessarily represent the official views of any of the funding agencies.

\section*{Diversity Statement}
Recent work in several fields of science has identified a bias in citation practices such that papers from women and other minority scholars are under-cited relative to the number of such papers in the field \cite{mitchell2013gendered,dion2018gendered,caplar2017quantitative, maliniak2013gender, Dworkin2020.01.03.894378}. Here we sought to proactively consider choosing references that reflect the diversity of the field in thought, form of contribution, gender, race, ethnicity, and other factors. First, we obtained the predicted gender of the first and last author of each reference by using databases that store the probability of a first name being carried by a woman \cite{Dworkin2020.01.03.894378,zhou_dale_2020_3672110}. By this measure (and excluding self-citations to the first and last authors of our current paper), our references contain 7.03\% woman(first)/woman(last), 14.47\% man/woman, 18.94\% woman/man, and 59.56\% man/man. This method is limited in that a) names, pronouns, and social media profiles used to construct the databases may not, in every case, be indicative of gender identity and b) it cannot account for intersex, non-binary, or transgender people. Second, we obtained predicted racial/ethnic category of the first and last author of each reference by databases that store the probability of a first and last name being carried by an author of color \cite{ambekar2009name, sood2018predicting}. By this measure (and excluding self-citations), our references contain 9.66\% author of color (first)/author of color(last), 17.21\% white author/author of color, 15.41\% author of color/white author, and 57.72\% white author/white author. This method is limited in that a) names and Florida Voter Data to make the predictions may not be indicative of racial/ethnic identity, and b) it cannot account for Indigenous and mixed-race authors, or those who may face differential biases due to the ambiguous racialization or ethnicization of their names.  We look forward to future work that could help us to better understand how to support equitable practices in science.

%\DB{How are we doing on citation diversity?} 

%\DB{Figures 2 and 3 would benefit from schematics explaining what you are doing and what you are testing.}

%
%
%\clearpage
\newpage
\bibliographystyle{plain}
\bibliography{library}

\end{document}